# PREDICTION OF CANCER MICROARRAY AND DNA METHYLATION DATA USING NON-NEGATIVE MATRIX FACTORIZATION

Parth Patel[1], Kalpdrum Passi[1$] and Chakresh Kumar Jain[2]

[1]Department of Mathematics and Computer Science,
Laurentian University, Sudbury, Ontario, Canada
[2]Department of Biotechnology, Jaypee Institute of Information Technology,
Noida, India

*ABSTRACT*

*Over the past few years, there has been a considerable spread of microarray technology in many biological patterns, particularly in those pertaining to cancer diseases like leukemia, prostate, colon cancer, etc. The primary bottleneck that one experiences in the proper understanding of such datasets lies in their dimensionality, and thus for an efficient and effective means of studying the same, a reduction in their dimension to a large extent is deemed necessary. This study is a bid to suggesting different algorithms and approaches for the reduction of dimensionality of such microarray datasets.This study exploits the matrix-like structure of such microarray data and uses a popular technique called Non-Negative Matrix Factorization (NMF) to reduce the dimensionality, primarily in the field of biological data. Classification accuracies are then compared for these algorithms.This technique gives an accuracy of 98%.*

## 1. INTRODUCTION

There has been an exponential growth in the amount and quality of biologically inspired data which are sourced from numerous experiments done across the world. If properly interpreted and analyzed, these data can be the key to solving complex problems related to healthcare. One important class of biological data used for analysis very widely is DNA microarray data, which is a commonly used technology for genome-wide expression profiling [1]. The microarray data is stored in the form of a matrix with each row representing a gene and columns representing samples, thus each element shows the expression level of a gene in a sample [2]. Gene expression is pivotal in the context of explaining most biological processes. Thus, any change within it can alter the normal working of a body in many ways and they are key to mutations [3]. Thus, studying microarray data from DNA can be a potential method for the identification of many ailments within human beings, which are otherwise hard to detect. However, due to the large size of these datasets, the complete analysis of microarray data is very complex [4]. This requires some initial pre-processing steps for reducing the dimension of the datasets without losing information.

Modern technology has made it possible to gather genetic expression data easier and cheaper from microarrays. One potential application for this technology is in identifying the presence and stage of complex diseases within an expression. Such an application is discussed further in this study.

This study, reflects upon the Non-Negative Matrix Factorization (NMF) technique which is a promising tool in cases of fields with only positive values and assess its effectiveness in the context of biological and specifically DNA microarray and methylation data. The results obtained are also compared with the Principal Component Analysis (PCA) algorithm to get relative estimates.

Motivation for the proposed work was two-fold, first, to test the use of Non-negative Matrix Factorization (NMF) as a feature selection method on microarray and methylation datasets and optimize the performance of the classifiers on reduced datasets. Although several feature selection methods, e.g. Principal Component Analysis (PCA), and others have been used in literature, there is limited use and testing of matrix factorization techniques in the domain of microarray and methylation datasets.

## 2. RELATED WORK

Extensive presence of DNA microarray data has resulted in undertaking lots of studies related to the analysis of these data, and their relation to several diseases, particularly cancer. The ability of DNA microarray data over other such data lies in the fact that this data can be used to track the level of expressions of thousands of genes. These methods have been widely used as a basis for the classification of cancer.

Golub et al. proposed in [5] a mechanism for the identification of new cancer classes and the consecutive assignment of tumors to known classes. In the study, the approach of gene expression monitoring from DNA microarray was used for cancer classification and was applied to an acute leukemia dataset for validation purposes.

Ramaswamy et al. [6] presented a study of 218 different tumor samples, which spanned across 14 different tumor types. The data consisted of more than 16000 genes and their expression levels were used. A support vector machine (SVM)-based algorithm was used for the training of the model. For reducing the dimensions of the large dataset, a variational filter was used which in its truest essence, excluded the genes which had marginal variability across different tumor samples. On testing and validating the classification model, an overall accuracy of 78% was obtained, which even though was not considered enough for confident predictions, but was rather taken as an indication for the applicability of such studies for the classification of tumor types in cancers leading to better treatment strategies.

Wang et al. [7] provided a study for the selection of genes from DNA microarray data for the classification of cancer using machine learning algorithms. However, as suggested by Koschmieder et al. in [4], due to the large size of such data, choosing only relevant genes as features or variables for the present context remained a problem. Wang et al. [12] thus systematically investigated several feature selection algorithms for the reduction of dimensionality. Using a mixture of these feature selections and several machine learning algorithms like decision trees, Naive Bayes etc, and testing them on datasets concerning acute leukemia and diffuse B-cell lymphoma, results that showed high confidence were obtained.

Sorlie et al. (2001) [8] performed a study for the classification of breast carcinomas tumours using genetic expression patterns which were derived from cDNA microarray experiments. The later part of the study also studied the clinical effects and implications of this by correlating the characteristics of the tumours to their respective outcomes. In the study, 85 different instances were studied which constituted of 78 different cancers. Hierarchical clustering was used for classification purposes. The obtained accuracy was about 75% for the entire dataset, and by using different sample sets for testing purposes.

Liu et al. [9] implemented an unsupervised classification method for the discovery of classes from microarray data. Their study was based on previous literature and the common application of microarray datasets which identified genes and classified them based on their expression levels, by assigning a weight to each individual gene. They modeled their unsupervised algorithm based on the Wilcoxon rank sum test (WT) for the identification and discovery of more than two classes from gene data.

Matrix Factorization (MF) is an unprecedented class of methods that provides a set of legal approaches to identify low-dimensional structures while retaining as much information as possible from the source data [22]. MF is also called matrix factorization, and the sequence problem is called discretization [22]. Mathematical and technical descriptions of the MF [19-20] methods, as well as microarray data [21] for their applications, are found in other reviews. Since the advent of sequencing technology, we have focused on the biological applications of MF technology and the interpretation of its results. [22] describes various MF methods used to analyze high-throughput data and compares the use of biological extracts from bulk and single-cell data. In our study we used specific samples and patterns for best visualization and accuracy.

Mohammad et al. [18] introduced a variational autoencoder (VAE) for unsupervised learning for dimensionality reduction in biomedical analysis of microarray datasets. VAE was compared with other dimensionality reduction methods such as PCA (principal components analysis), fastICA (independent components analysis), FA (feature analysis), NMF and LDA (latent Dirichlet allocation). Their results show an average accuracy of 85% with VAE and 70% with NMF for leukemia dataset whereas our experiments provide 98% accuracy with NMF. Similarly, for colon dataset, their results show an average accuracy of 68% with NMF and 88% with VAE whereas our experiments provide 90% accuracy with NMF.

## 2.1. Contributions

1. Two different types of high dimensional datasets were used, DNA Microarray Data (Leukemia dataset, prostate cancer dataset, colon cancer dataset) and DNA methylation Data (Oral cancer dataset, brain cancer dataset)
2. Two different feature extraction methods, NMF and PCA were applied for dimensionality reduction on datasets with small sample size and high dimensionality using different classification techniques (Random Forest, SVM, K-nearest neighbor, artificial neural networks).
3. All the parameters were tested for different implemntations of NMF and classifiers and the best parameters were used for selecting the appropriate NMF algorithm.
4. Computing time Vs. number of iterations were compared for different algorithms using GPU and CPU for the best performance.
5. Results demonstrate that NMF provides the optimum features for a reduced dimensionality and gives best accuracy in predicting the cancer using various classifiers.

## 3. MATERIAL AND METHODS

### 3.1. Datasets

As a part of this study, five datasets relating to cancer were analyzed of which three are microarray datasets and two are methylation datasets. The aim of using methylation dataset is to ascertain the impact of DNA methylation on cancer development, particularly in the case of central nervous system tumors. The Prostate Cancer dataset contains a total of 102 samples and 2135 genes, out of which 52 expression patterns were tumor prostate specimens and 50 were normal specimens [6]. The second dataset used as a part of this study is a Leukemia microarray dataset [5]. This dataset contains a total of 47 samples, which are all from acute leukemia patients. All the samples are either acute lymphoblastic leukemia (ALL) type or of the acute myelogenous leukemia (AML) type. The third dataset used is a Colon Cancer microarray Dataset (Alon et al. 1999) [11]. This dataset contains a total of 62 samples, out of which 40 are tumor samples and the remaining 22 are from normal colon tissue samples. The dataset contained the expression samples for more than 2000 genes having the highest minimal intensity for a total of 62 tissues. The ordering of the genes was placed in decreasing order of their minimal intensities. The fourth and fifth dataset are the brain cancer and the oral cancer dataset with extensive DNA methylation observed. (Capper et al. 2018) [10]. the cancer methylation dataset is a combination of two types of information which are the acquired DNA methylation changes and the characteristics of cell origin. It is observed that such DNA methylation profiling is highly potent for the sub-classification of central nervous system tumors even in cases of poor-quality samples. The datasets are extensive with over 180 columns and 21000 rows denoting different patient cases. Classification is done using 1 for positive and 0 for negative instances.

### 3.2. Feature Selection

In the field of statistics and machine learning, feature reduction refers to the procedure for reducing the number of explanatory (independent) columns from the data under consideration. It is important in order to reduce the time and computational complexity of the model while also improving the robustness of the dataset by removal of correlated variables. In some cases, such as in the course of this study, it also leads to better visualization of the data.

Some of the popular feature reduction techniques include Principal Component Analysis (PCA), Non-negative Matrix Factorization (NMF), Linear Discriminant Analysis (LDA), autoencoders, etc.

In this study, we analyze Non-negative Matrix Factorization (NMF) technique for feature selection and test the effectiveness in the context of biological and specifically DNA microarray and methylation data and compare our results with the PCA algorithm to get relative estimates.

#### 3.2.1. Non-negative Matrix Factorization (NMF)

Kossenkov et al. (2010) [2] suggests the use of several matrix factorization methods for the same problem. The method presented in the paper is a Non-negative Matrix Factorization (NMF) technique which has been used for dimensionality reduction in numerous cases (Wang et al. 2013) [5].

For a random vector X with (m x n) dimensions, the aim of NMF is to try to express this vector X in terms of a basis matrix U (m x l) dimension and a coefficient matrix V (l x n) dimension, i.e.:

*X ≈ UV*

The initial condition being L << min (m, n).

Here U and V are m x l and l x n dimensional matrices with positive values. U is known as the basis matrix while V is the coefficient matrix. The idea behind this algorithm is to obtain values of U and V such that the following function is at its local minima:

$$min[D(X, UV) \, + \, R(U, V)]$$

where D: distance cost function, R: regularization function.

Results show that the obtained matrix U has a high dimension reduction from X.

On the application of the NMF algorithms, similar to other dimensionality reduction algorithms, a large number of variables are clustered together. In the case of NMF, expression data from the genes are reduced to form a small number of meta-genes (Brunet et al. 2004) [13]. For the matrix U, each column represents a metagene and the values give the contribution for the gene towards that metagene. The working of an NMF algorithm on a microarray dataset is shown in the figure below, with each pixel indicating the expression values (shown in the form of the intensity of colors).

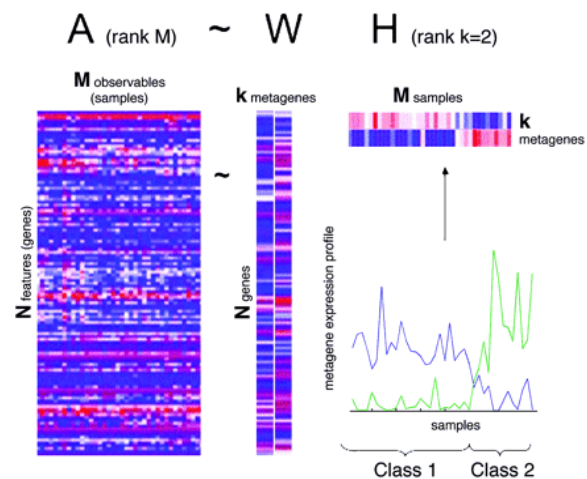

Figure 1. Image showing how an NMF algorithm is used to get a basis matrix of rank (2) [13]

After subsequent NMF based dimensionality reduction is done, the obtained reduced matrix is expected to contain the same information as the original matrix. In order to check the validity of the above assumption, classification algorithms are applied to the reduced matrix and subsequently, their accuracies were measured. Another popular dimensionality reduction technique which is the Principal Component Analysis (PCA) is used and then compared with the proposed NMF based models. It must be emphasized that conventional NMF based algorithms, even though very accurate are highly resource intensive when used for large datasets. This study, therefore, also makes use of certain GPU algorithms in order to effectively evaluate the same, making training time much more feasible.

There are different implementations of the NMF which are categorized based on the choice for the loss function (D) and the regularization function (R). The implementations used in this study are listed below.

- Non-smooth NMF (nsNMF) [14]
- Kullback Leibler Method (KL) [13]
- Frobenius [15]
- Offset [15]
- Multiplicative Update Algorithm (MU) [15]
- Alternating Least Square (ALS) [16]
- Alternating Constrained Least Square (ACLS) [17]
- Alternating Hoyer Constrained Least Square (AHCLS) [17]
- Gradient Descent Constrained Least Square (GDCLS) [15]

### 3.2.2. Principal Component Analysis (PCA)

PCA is a widely used technique in the field of data analysis, for orthogonal transformation-based feature reduction on high-dimensional data. Using PCA, a reduced number of orthogonal variables can be obtained which explain the maximum variation within the data. Using a reduced number of variables helps in significant reduction in computational cost and runtimes while still containing a high amount of information as within the original data.

It is a widely used tool throughout the field of analytics for feature reduction before predictive modelling. The steps involved in the computation of a PCA algorithm is shown below:

Let '$X$' be the initial matrix of dimension (m $x$ n), where m = number of rows, and n is the number of columns.

The first step is to linearly transform the matrix $X$ into a matrix $B$ such that,

$$B = Z * X$$

where, $Z$ is a matrix of order (m $x$ m).

The second step is to normalize the data. In order to normalize, the mean for the data is computed and normalization is done by subtracting off the mean for finding out the principal components. The equations are shown below:

$$M(m) = 1/N \sum_{n=1}^{N} X[m,n], X' = X - M$$

The next step involves computing the covariance matrix of X, which is computed as below

$$C_X = X * \frac{X^T}{(n-1)}$$

In the covariance matrix, all diagonal elements represent the variance while all non-diagonal elements represent co-variances.
The covariance equation for B is shown below

$$C_B = B * \frac{B^T}{(n-1)} = \frac{(ZA)(ZA)^T}{(n-1)} = \frac{(ZA)(A^T*Z^T)}{(n-1)} = \frac{ZY*Z^T}{(n-1)}$$

where $Y = A * A^T$, and of dimension (m x m),

Now, Y can be further expressed in the form, Y = EDE, where E is an orthogonal matrix whose columns represent the eigenvalues of Y, while D is a diagonal matrix with the eigenvalues as its entries.

If $Z = E^T$, the value of covariance for B becomes,

$$C_B = \frac{ZY*Z^T}{(n-1)} = \frac{E^T(ED*E^T)E}{(n-1)} = \frac{D}{(n-1)}$$

The eigenvalues in this case are arranged in descending order, thus the most important component comes first and so on.

Thus, from the transformed matrix 'B', only a subset of features can be taken which preserve a larger share of the variance within the data.

### 3.2.3. Parameter Selection

Table 1 summarizes the different parameters which have been used for different algorithms.
Table 1: Parameters

| Algorithm | Parameters |
|---|---|
| Mu | Rank = 5 |
| GDCLS | Rank = 5, λ= 0.1 |
| ALS | Rank = 5 |
| ACLS | Rank = 5, λH=0.1, λW=0.1 |
| AHCLS | Rank = 5, λH=0.1, λW=0.1, αH=0.5, αW=0.5 |
| PCA | Number of features = varying between 1 to 100. row.W = 1, col.W = 1 |
| Random Forest | nTrees = 500, cutoff = ½, nodesize = 1 |
| SVM | coef = 0, cost = 1, nu = 0.5, tolerance = 0.001 |

### 3.2.4. Runtime Computation using CPU and GPU

There exist many different algorithms for implementation of NMF using R. These algorithms can be broadly classified as those which are implemented using the CPU architecture of the system and those which are implemented using GPU architectures. The CPU codes are implemented using the standard 'NMF' package within R while the GPU codes are implemented using a modification of the 'NMF' package known as the 'NMFGPU4R" package which uses multicore options from GPUs to massively parallelize the implementation of the algorithms.

The different algorithms are first run on the three datasets and their run times are noted respectively. The runtime is defined as the amount of time (in seconds) taken by the computer to compute that respective algorithm. A higher runtime generally means high complexity and should be avoided, as such algorithms generally don't scale very well. During each of the implementation, we chose three clusters within the output dataset in order to provide uniformity.

For the algorithms executed on CPUs, the obtained run times are given in Table 2.

Table 2. CPU times (seconds) of different NMF algorithms on the cancer datasets

| Method | Prostrate | Colon | Leukemia |
|--------|-----------|-------|----------|
| NsNMF | 90.43 | 169.19 | 157.1 |
| KL | 9.85 | 8.29 | 6.08 |
| Frobenius | 3.86 | 4.16 | 3.09 |
| Offset | 103.63 | 164.62 | 160.50 |

As can be seen clearly for the three datasets, the **_Offset_** method takes the highest runtime of over 100s for prostate cancer data and over 160s for colon cancer and leukemia data, closely preceded by the nsNMF method which is just under 160s. The Frobenius and the KL methods take significantly lesser times (under 10s) than the other two methods. These algorithms could not be computed on the methylation datasets using CPUs as these are very high dimensional (over 10000 rows) datasets.

For the algorithms executed on GPUs, the obtained run times are given in Table 3.

Table 3. GPU times (seconds) of different NMF algorithms on the cancer datasets

| Method | Microarray dataset | Brain Cancer | Oral Cancer |
|--------|-------------------|--------------|-------------|
| Mu | 0.66 | 5.01 | 4.18 |
| ALS | 0.77 | 1.823 | 1.20 |
| GDCLS | 1.44 | 0.34 | 0.309 |
| NSNMF | 0.96 | 4.334 | 4.114 |
| ACLS | 0.58 | 1.287 | 1.19 |
| AHCLS | 0.51 | 1.11 | 0.67 |

As can be seen from Table 3, the AHCLS and the ACLS are the quickest to converge with runtimes of 0.51s and 0.58s respectively. However, the brain and oral cancer datasets had different outcomes in this regard. The GDCLS method was the quickest to converge for brain and oral cancer datasets with a runtime of 0.34s and 0.309s respectively.

### 3.3. Classification Algorithms

The classification algorithms used in this study include random forest, support vector machine, neural networks and k-nearest neighbors. These are briefly explained.

#### 3.3.1. Random Forest (RF)

The random forest algorithm is a supervised learning algorithm which is widely used as an ensemble model for classification and regression tasks. Classification involves training a model with a set of independent variables in order to output a dependent variable into certain pre-defined factors. The random forest algorithm is an ensemble of decision trees. The output is a mean of the output for the different decision trees. The training algorithm from random forest is based on bootstrap aggregating to tree classifiers. Generally, for a given training sample $(x_1, x_2, \ldots x_n)$ along with their respective response variables $(y_1, y_2, \ldots y_n)$, for $b = 1 \ldots B$: random samples are selected with replacement from the n training examples and the classifiers are then trained on them using trees $f_b$ .

The equation for prediction for out-of-sample data in case of a random classifier is shown below:

$$p = \frac{1}{B}\sum_{b=1}^{B} \mathcal{F}(x)$$

### 3.3.2.  Support Vector Machine (SVM)

The Support Vector Machine algorithm is a very commonly used algorithm for classification and regression. The models is a non-probabilistic kind where each individual data point is represented in an n-dimensional space where n refers to the number of independent features. Classification is done by fitting an (n-1) dimensional plane within the space and on the basis of the position of a particular data point with respect to the plane. The confidence of prediction is determined by the distance of the position of the point with respect to the dividing plane.

Let $(x_1, y_1), (x_2, y_2)\ldots\ldots\ldots(x_2, y_2)$ be a set of n points where $x$ represents a p-dimensional vector and $y$ is an indicator variable for the class which takes the value 0 or 1. The goal of SVM is to find a dividing hyperplane that divides the entire n-points based on their indicator variable value.  Such a hyperplane can be represented as

$$\underline{w}.\underline{x} - b = 0$$

where, $w$ is the normal vector and, b is a parameter which determines the offset for the hyperplane.

### 3.3.3.  Neural Network (NN)

Neural Networks comprise of a collection of different algorithms and paradigms which resemble the networks within the human brain in the context that they take input and process it through different layers and nodes. Inputs are passed on as vectors and the neural networks perform operations which help us cluster or regress this data.

The primary constituent of a neural network is a set of different layers Each layer is further made up of nodes. This entire structure is modelled based on the neural structure of a human brain. With each node is associated a set of weights and coefficients which largely determine the amplification of the amount of signal of data which passes through it. The final layer contains a single node which sums up the values from all the nodes in its preceding layer and outputs a single value indicating the label (classification) or the value of the output (regression).

### 3.3.4.  K-nearest Neighbor (KNN)

The K-nearest Neighbor (KNN) model is a non-parametric model which is commonly used in the field of machine learning for the purpose of classification or regression tasks.

In this method, the input is a vector of some 'k' closest training examples which exist in the hyperspace and based on the value of these k-nearest neighbors, the classification for the object is done based on a majority voting about the classes of its neighbors.

## 4. RESULTS

In order to evaluate the performance of classification algorithms on the NMF and PCA reduced data four classifiers were trained, random forest (RF), support vector machine (SVM), neural network (ANN) and k-nearest neighbor (KNN). The classifiers were tested for accuracy and area under the ROC curve (AUC) performance metrics. 10-fold cross validation was used to avoid overfitting and different number of features were selected using NMF and PCA methods. Figure 2 shows the accuracy of the four classifiers with NMF and PCA methods on the leukemia dataset.

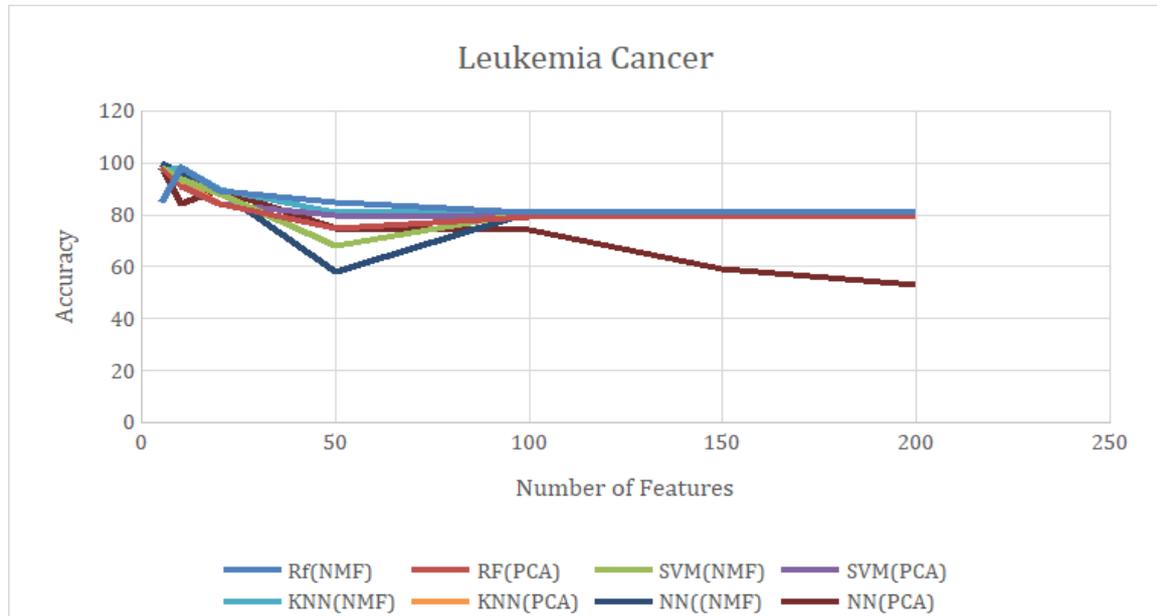

Figure 2. Accuracy of classifiers with different number of features for leukemia dataset

It is evident from Figure 2 that the accuracies obtained for lower number of features (10) are the most accurate with about 98% accuracy obtained for most classifiers in case of NMF and a very high AUC (~0.97). For PCA reduced matrices, the highest accuracy is also in the range 0f 98% but for lessernumber of features (5).

Figure 3 shows the accuracy of the four classifiers with NMF and PCA methods on the prostate dataset.

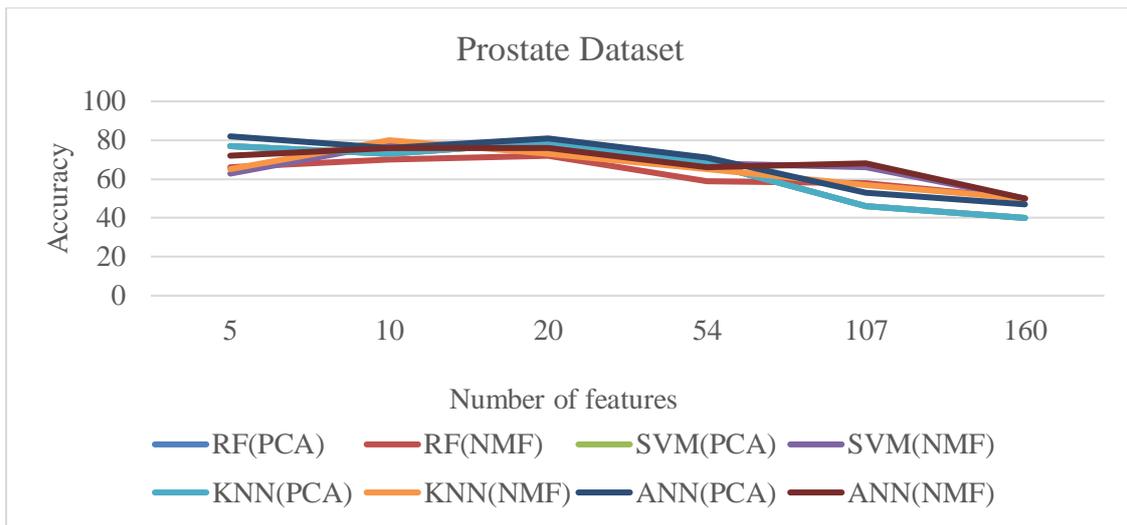

Figure 3. Accuracy of classifiers with different number of features for prostate dataset

It is evident from Figure 3 that the accuracy generally decreases as the number (percentage) of features increase. For lesser number of features (10), the accuracy is in the range of 70-80%. However, as the features increase, the accuracy falls to about 50%. The AUC values also show a similar trend.

Figure 4 shows the accuracy of the four classifiers with NMF and PCA methods on the colon dataset.

In the case of colon dataset, the SVM classifier gives an accuracy in the range of about 87% for lower number of features (10). As in the case of other datasets, there is a general decrease in the accuracy as the number of features increases.

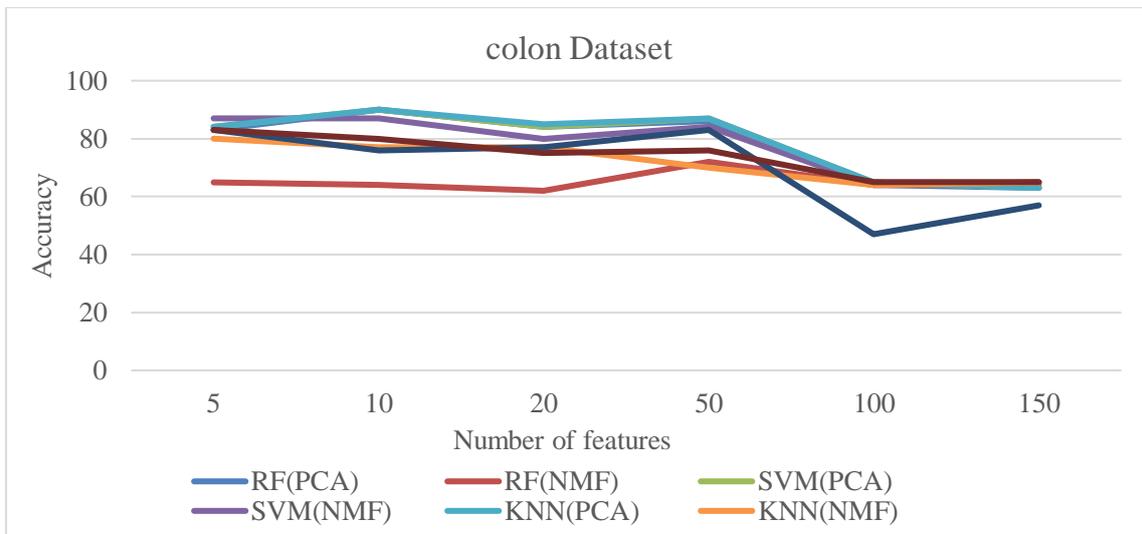

Figure 4. Accuracy of classifiers with different number of features for colon dataset

Figure 5 shows the accuracy of the four classifiers with NMF and PCA methods on the oral cancer dataset.

The Oral Cancer dataset has a general low accuracy across all classifiers (~65%) for all the different combinations of number of features. However, the AUC values are high (~0.95).

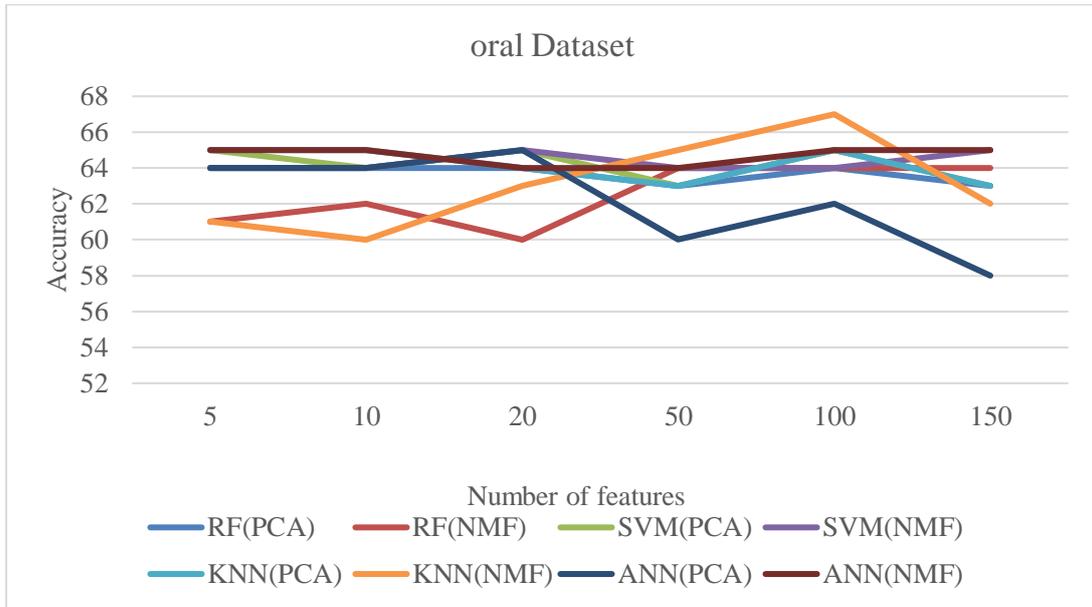

Figure 5. Accuracy of classifiers with different number of features for oral cancer dataset

Figure 6 shows the accuracy of the four classifiers with NMF and PCA methods on the brain cancer dataset.

Both the NMF and the PCA algorithms perform similarly in this case with the highest accuracy being 92% and 95% respectively. The ANN algorithm shows a consistent performance of over 90% in this case.

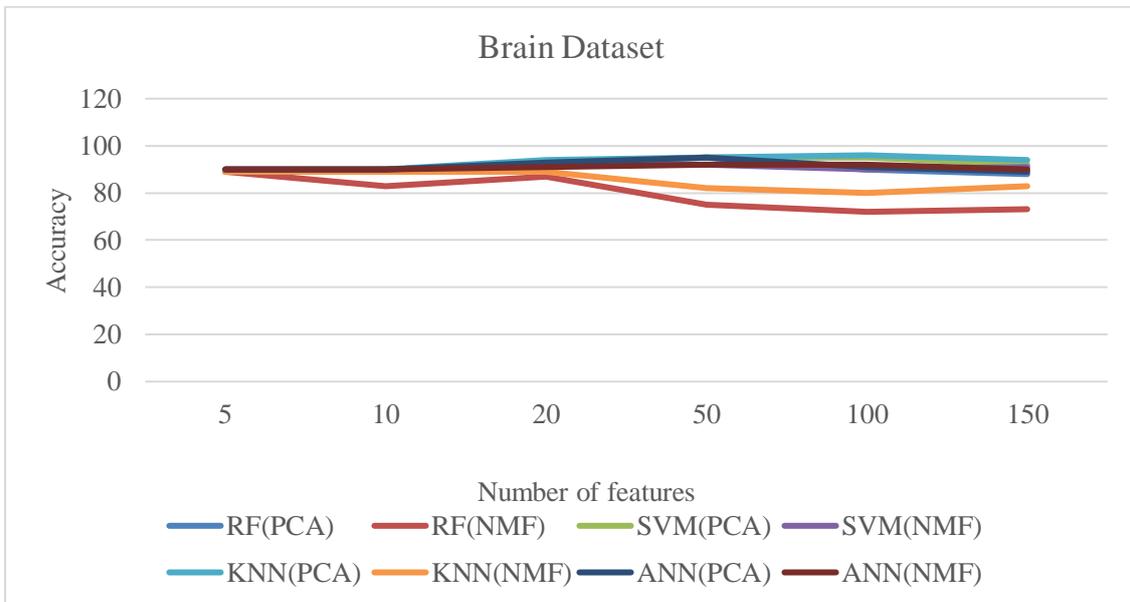

Figure 6. Accuracy of classifiers with different number of features for brain cancer dataset

For the second phase of our analysis on the classification algorithms, different training-to-testing ratios were taken to find the best accuracy and AUC values.

For the leukemia dataset, it is observed that for lower number of features, the ANN method is the best in terms of accuracy (~99%) as opposed to over 80% shown by other classifiers. However, as the number of features increase the accuracy for all the cases decreases. In the case for the Prostate Cancer dataset, accuracies in the range of 70-90% are obtained throughout the different combinations. The highest accuracy is obtained for the training-to-testing ratio 50:50 while using ANN for a small number of features (5). The accuracy in this case is close to 100%. For the colon cancer dataset, it is seen that the raw data has the best accuracy for the RF and SVM models while the highest accuracy is obtained for the ANN model (95%). The brain cancer dataset is the first of methylation datasets. It is seen that the performance of the two algorithms is consistent over the range of split percentages i.e. (~90%), for the Random Forest classifier with reduced features (5). The ANN also performs with a similar accuracy along with the KNN model (accuracy ~89%). A lower accuracy is obtained for the oral cancer dataset across all combinations. The overall range for the accuracy is between 60-80% barring a few exceptions. The ANN is the most consistent classifier here with classification accuracies above 80% in all other cases except when the number of reduced features is the highest (5).

Table 4 shows the summary results of the best accuracies for each classifier and dataset for NMF and PCA algorithms. After performing all the analysis, the following points become pertinent:

1.  Random Forest and KNN classifiers have an accuracy of 98% after using just 5 and 10 of the number features in the leukemia dataset for the NMF and PCA reduced datasets respectively.
2.  SVM algorithm also gives 98% accuracy for the leukemia dataset with only 5 features.
3.  ANN algorithm gives 100% accuracy with 5 features used with NMF. However, the accuracy for the other datasets are lower with the Brain Cancer dataset giving over 90% accuracy.
4.  For the colon cancer dataset, PCA generally performs better than the NMF algorithm as RF, SVM and KNN have about 90% accuracy with PCA, however it is about 80% with NMF.
5.  The PCA+RF and PCA+ANN combination in case of brain cancer data gives 95% accuracy with 50 features.
6.  The oral cancer dataset has the lowest accuracies, with the highest being 69% using the PCA+SVM classifier on the Raw Data.

Table 4. Summary results of best accuracy and number of features

| Classifier | Dataset | NMF | | PCA | |
|---|---|---|---|---|---|
| | | Acc. | # of features | Acc. | # of features |
| **RF** | Leukemia | 98 | 10 | 98 | 5 |
| | Prostate Cancer | 72 | 20 | 77 | 5 |
| | Colon Cancer | 74 | 10 | 90 | 20 |
| | Brain Cancer | 89 | 5 | 95 | 50 |
| | Oral Cancer | 65 | 180 | 65 | 180 |
| **SVM** | Leukemia | 98 | 5 | 98 | 5 |
| | Prostate Cancer | 78 | 150 | 79 | 20 |
| | Colon Cancer | 87 | 5 | 90 | 10 |
| | Brain Cancer | 92 | 20 | 95 | 50 |
| | Oral Cancer | 66 | 180 | 69 | 180 |
| **KNN** | Leukemia | 98 | 10 | 98 | 5 |
| | Prostate Cancer | 80 | 10 | 79 | 20 |
| | Colon Cancer | 80 | 5 | 90 | 10 |
| | Brain Cancer | 89 | 10 | 96 | 100 |
| | Oral Cancer | 67 | 100 | 65 | 10 |
| **ANN** | Leukemia | 100 | 5 | 98 | 5 |
| | Prostate Cancer | 76 | 20 | 81 | 20 |
| | Colon Cancer | 83 | 5 | 83 | 5 |
| | Brain Cancer | 92 | 100 | 95 | 50 |
| | Oral Cancer | 65 | 150 | 65 | 180 |

# 5. CONCLUSIONS AND FUTURE WORK

In this study, NMF has been used on several microarray datasets in order to obtain dimensionality reduction in the feature sets. After application of NMF algorithms, the number of variables from the gene expression data using microarray profiles went down from about 1000 to a handful. On reduction, the obtained datasets were easier to read both visually and through heat maps and plots. An optimum number of reduced features were obtained and for this, individual heap maps and coefficient-maps were plotted. Unlike previous methods, where reducing the number of variables required extensive study on the nature of the datasets, using this approach, the same could be done through numerical computations on the given datasets.

This study is of immense biological significance. Its significance lies in detecting and identifying genes for differentiating cancer diseases and non-cancer diseases so that a proper tailor-made treatment can be initiated. It also recognizes specific biomarkers of cancer which can be further analyzed. It is essential to identify genes which play a role in the development of a cancer as the gene expressions of patients suffering from cancer are different from those of healthy patients. It also makes analyzing data easier for using machine learning techniques.

Throughout the course of the study there are two major drawbacks which are elucidated below. Due to limited computational resources, the entire set of NMF algorithms couldn't be run on every dataset, particularly the methylation ones which have higher dimensions, leading to certain imperfections in their analysis. This also resulted in our inability to perform semi-NMF calculations as well as residual analysis plots for these datasets. The lack of computational power was again the major reason for keeping the subset feature numbers restricted to 50. As any number of features higher than that would exponentially increase the training time for the classifiers.

As discussed above, NMF has potential use for feature reduction of high dimensional data, particularly in the context of microarray data. A significant work can be done in the direction of classifying different subgroups of a particular disease by first using NMF to reduce high dimensional microarray data. This not only simplifies the process but also provides a faster and more robust model of diagnosis.

Use of NMF for methylation datasets is limited in the context that a lower accuracy is observed for these cases. Further studies should revolve around improving the scenario of the same.

**AUTHORS**

**Parth Patel** graduated from Laurentian University, Ontario, Canada with Master's degree in Computational Sciences. He pursued his bachelor's degree in Computer Engineering from the Gujarat Technological University, Gujarat, India. His main areas of interest are Machine Learning Techniques, Data Analytics and Data Optimization Techniques. He completed his M.Sc. thesis in Prediction of Cancer Microarray and DNA Methylation Data Using Non-Negative Matrix Factorization.

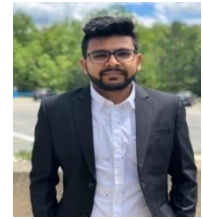

**Kalpdrum Passi** received his Ph.D. in Parallel Numerical Algorithms from Indian Institute of Technology, Delhi, India in 1993. He is an Associate Professor, Department of Mathematics & Computer Science, at Laurentian University, Ontario, Canada. He has published many papers on Parallel Numerical Algorithms in international journals and conferences. He has collaborative work with faculty in Canada and US and the work was tested on the CRAY XMP's and CRAY YMP's. He transitioned his research to web technology, and more recently has been involved in machine learning and data mining applications in bioinformatics, social media and other data science areas. He obtained funding from NSERC and Laurentian University for his research. He is a member of the ACM and IEEE Computer Society.

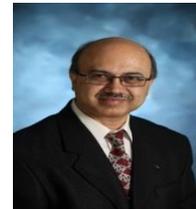

**Chakresh Kumar Jain** received his Ph.D. in the field of bioinformatics from Jiwaji University, Gwalior, India, focusing on computational designing of non-coding RNAs using machine learning methods. He is an Assistant Professor, Department of Biotechnology, Jaypee Institute of Information Technology, Noida, India. He is CSIR-UGC-NET [LS] qualified and member of International Association of Engineers (IAENG) and life member of IETE, New Delhi, India. His research interests include development of computational algorithms for quantification of biological features to decipher the complex biological phenomenon and diseases such as cancer and neurodegenerative diseases apart from drug target identification and mutational analysis for revealing the antibiotic resistance across the microbes through computer based docking, molecular modelling and dynamics, non-coding RNAs identification, machine learning, data analytics, and systemsbiology based approaches.

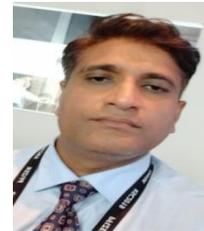